# III-V-on-silicon triple-junction based on the heterojunction bipolar transistor solar cell concept

Elisa Antolín
*Universidad Politécnica de Madrid*
*Instituto de Energía Solar*
Madrid, Spain.
elisa.antolin@upm.es
ORCID: 0000-0002-5220-2849

Marius H. Zehender
*Universidad Politécnica de Madrid*
*Instituto de Energía Solar*
Madrid, Spain.
marius.zehender@upm.es
ORCID: 0000-0002-2263-4560

Simon A. Svatek
*Universidad Politécnica de Madrid*
*Instituto de Energía Solar*
Madrid, Spain.
simon.svatek@upm.es
ORCID: 0000-0002-8104-1888

Pablo García-Linares
*Universidad Politécnica de Madrid. Solar*
*Instituto de Energía Solar*
Madrid, Spain.
p.garcia-linares@upm.es

Antonio Martí
*Universidad Politécnica de Madrid. Solar*
*Instituto de Energía Solar*
Madrid, Spain.
antonio.marti@upm.es

*Abstract*— We propose a new triple-junction solar cell structure composed of a III-V heterojunction bipolar transistor solar cell (HBTSC) stacked on top of, and series-connected to, a Si solar cell (III-V-HBTSC-on-Si). The HBTSC is a novel three-terminal device, whose viability has been recently experimentally demonstrated. It has the theoretical efficiency limit of an independently-connected double-junction solar cell. Here, we perform detailed balance efficiency limit calculations under one-sun illumination that show that the absolute efficiency limit of a III-V-HBTSC-on-Si device is the same as for the conventional current-matched III-V-on-Si triple-junction (47% assuming black-body spectrum, 49% with AM1.5G). However, the range of band-gap energies for which the efficiency limit is above 40% is much wider in the III-V-HBTSC-on-Si stack case. From a technological point of view, the lattice-matched GaInP/GaAs combination is particularly interesting, which has an AM1.5G efficiency limit of 47% with the HBTSC-on-Si structure and 39% if the current-matched III-V-on-Si triple junction is considered. Moreover, we show that interconnecting the terminals of the HBTSC to achieve a two-terminal GaInP/GaAs-HBTSC-on-Si device only reduces the efficiency limit by three points, to 43%. As a result, the GaInP/GaAs-HBTSC-on-Si solar cell becomes a promising device for two-terminal, high-efficiency one-sun operation. For it to also be cost-effective, low-cost technologies must be applied to the III-V material growth, such as high-throughput epitaxy or sequential growth.

## I. Introduction

A multi-junction solar cell (MJSC) consists of a number of sub-cells made of different materials, stacked on top of each other and ordered so that the band-gap energy of each cell is smaller than the band-gap energy of the sub-cell immediately above it. In this arrangement, sub-cells with higher band-gap energy can absorb photons of higher energy and deliver correspondingly higher voltages. Thus, the total output of the cell can be greatly enhanced compared to the situation where a single-gap solar cell would absorb all photon energy ranges. In practice, this has been demonstrated for up to six junctions and a six-junction solar cell produced last year by NREL [1] holds the current absolute efficiency record of photovoltaic (PV) technology (47.1% at 143 suns and 39.2% at one sun).

Each sub-cell in a multi-junction stack could allocate two electrical terminals and, in principle, those terminals could be interconnected at will between sub-cells. In practice, most MJSCs have all their sub-cells connected in series, i. e. current-matched. This configuration has been very successful because the resulting MJSCs are two-terminal devices and their inter-connection in a module is straightforward.

However, the current-matching constraint in MJSCs also poses some problems, such as strong spectral sensitivity and a very restricted range of band-gap energies that lead to optimum performance. In the last years it has become apparent that those factors, and not only the conversion efficiency, have great influence on the final electricity production potential of a PV technology. This has motivated interest in device architectures with three or more terminals [2], which can produce more power under variable spectrum or temperature changes.

Another important change in the MJSCs technological roadmap has to do with cost. The cost of silicon cells has become so low compared to all competing PV technologies, that MJSCs built on a silicon bottom cell and operating under one-sun seem to be the most cost-effective option to increase the efficiency of commercial devices, at least in the short term. The possible materials to combine with silicon include II-VI compounds, perovskites, organic semiconductors, and the highly efficient III-V compounds.

## II. Proposed solar cell structure

In this work, we propose a new type of MJSC containing a total of three junctions, with the bottom one being made of silicon. This device is depicted in Fig. 1a. The two upper junctions are

This work has been funded by the Spanish Science Ministry under Grant RTI2018-096937-B-C21, by the European Union's Horizon 2020 research and innovation program under grant agreement No. 787289, and by Fundación Iberdrola within the ConCEPT II Project. M.H.Z. is grateful to Universidad Politécnica de Madrid for funding through the Predoctoral Grant Programme. E.A. acknowledges a Ramón y Cajal Fellowship (RYC-2015-18539) and S.A.S acknowledges a Juan de la Cierva Fellowship (FJC2018-036517-I), both funded by the Spanish Science Ministry.

not two separate sub-cells as in conventional MJSCs. Instead, they are integrated in a heterojunction bipolar transistor solar cell (HBTSC) [3]. The HBTSC is a double-junction device which has an extremely simple structure based on three layers, called emitter, base and collector, and three terminals that enable the independent operation of the two junctions. In the HBTSC-on-Si device the back contact of the HBTSC is connected to the top contact of the Si cell. This results in a three-terminal device. It is also possible to interconnect the terminals to reduce the number to two, as we will show later.

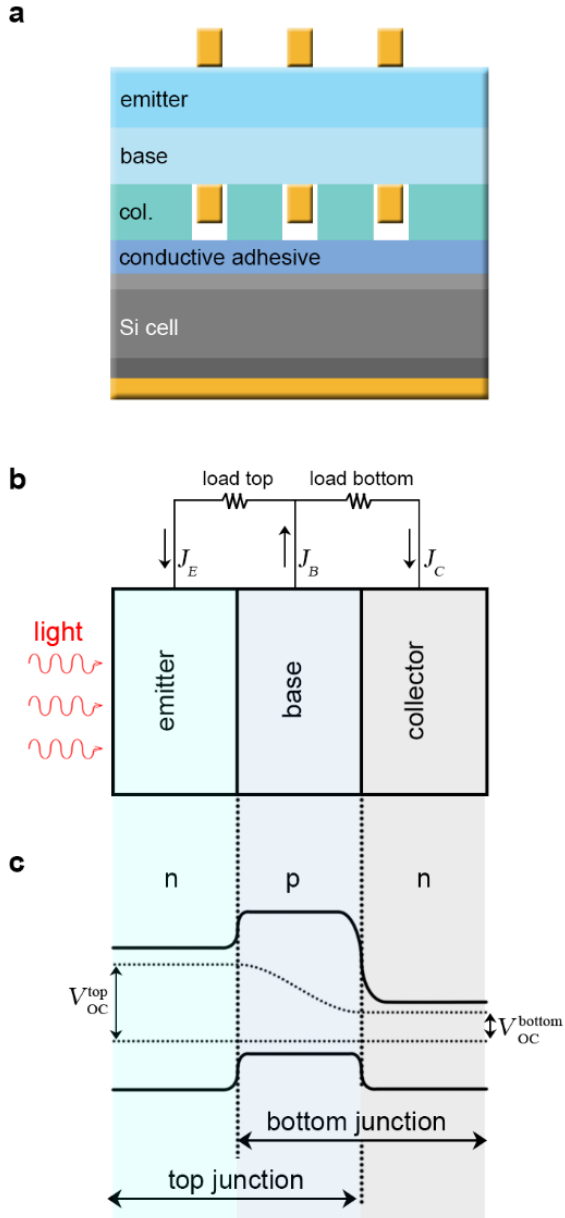

Fig. 1: (a) Proposed device: III-V-on-silicon triple-junction using an HBTSC as top and middle junctions and a Si cell as bottom junction. (b)Structure of an *npn* HBTSC showing the three-terminals and the external circuits for independent current extraction. (c) Band diagram of an *npn* HBTSC in open circuit. Dashed lines represent the electron and hole quasi-Fermi levels.

The concept of HBTSC is material-agnostic, but in this paper, we will focus on the case where the HBTSC is made of III-V semiconductors. For the III-V-HBTSC-on-Si device to be cost-effective, the III-V HBTSC has to be fabricated with low-cost techniques, transferred to the top of the silicon cell and glued with a low-cost transparent conductive adhesive [4], [5]. If other materials are chosen, potentially cheaper fabrication techniques could be employed, possibly at the cost of efficiency. With II-VI compounds, for example, the HBTSC could be sputtered directly onto the top of the silicon cell. The trade-off between efficiency and cost will eventually determine which option has a greater potential.

The basic layer structure of an HBTSC is shown in Fig. 1b [3]. It resembles a heterojunction bipolar transistor and, as in the case of the transistor, it can be either *npn* or *pnp*. The top junction is formed between the emitter and base and the bottom junction between base and collector. To work as a MJSC, the emitter has a large, and the collector a small, band-gap energy. The base can have the same or a greater band-gap energy as the emitter.

The key of the HBTSC theoretical model is that it is possible to achieve a double-junction characteristic if the minority carrier quasi-Fermi level bends across the base layer (see Fig.1c) enabling that the top-junction open-circuit voltage ($V_{OC}^{top}$) is larger than the bottom-junction open-circuit voltage ($V_{OC}^{bot}$) [3]. This has been recently demonstrated in a GaInP/GaAs HBTSC fabricated by our group in collaboration with NREL [6], [7]. In that proof-of-concept device the base layer was 800 nm thick. AM1.5G $V_{OC}$s of 0.95 and 1.33 V were recorded at the bottom and top junctions, respectively, and the overall efficiency reached 28%. More recently, we have made the first inverted GaInP/GaAs HBTSC, transferring it successfully onto a Si carrier [8], opening the path to the practical implementation of a III-V-HBTSC-on-Si triple-junction device.

### III. DETAILED BALANCE EFFICIENCY CALCULATIONS

Fig. 2 shows three possible III-V-on-Si triple-junction devices, which can be fabricated by detaching the III-V structure from its epitaxial substrate and attaching it to a Si solar cell using a transparent conductive adhesive. Fig. 2a shows the conventional current-matched triple-junction solar cell (3JSC) with Si at the bottom. Fig 2b shows the III-V-HBTSC-on-Si structure proposed in this work in its three-terminal version. Fig. 2c depicts a similar III-V-HBTSC-on-Si triple-junction device in which the emitter contact has been short-circuited with the Si cell bottom contact, resulting in a two-terminal device. It is interesting to note that b and c are 3JSCs with no tunnel junctions.

We have depicted under each device the corresponding equivalent circuit model. The circuits help understanding the detailed balance efficiency calculations because they illustrate the constraints between sub-cells. In device b there is only one constraint, the series connection between the collector and the Si cell, whereas a and c are subject to two constrains. The two-terminal current-matched device (a) includes another cell in series and the two-terminal device based on the HBTSC concept (c) contains both a parallel constraint and a series constraint.

In the equivalent circuits we have assumed that the HBTSC behaves as two opposed junctions. This is true if the cell design prevents carrier injection from the emitter into the base by implementing a large band-gap energy and/or a high doping level in the base [9]. If carrier injection was a possibility, some of the injected carriers could reach the base-collector junction and a more complex equivalent circuit would be required to reflect that fact, such as the Ebers-Moll equivalent circuit (widely used for bipolar transistors). It can be demonstrated that carrier injection from the emitter into the base introduces an entropy loss term in the theoretical model of the HBTSC [3]. This term appears because, in the presence of carrier injection, the bending of the minority quasi-Fermi level in the base (required to achieve $V_{OC}^{top} > V_{OC}^{bot}$) results in ohmic losses. Suppressing this carrier injection maximizes the theoretical efficiency of the HBTSC, making it equal to the efficiency of two opposed junctions with independent voltages. Therefore, we will only consider the case of zero minority carrier injection because it does not only lead to the simplest equivalent circuit, but also to the highest efficiency. In [9] it is demonstrated that carrier injection can be eliminated using realistic III-V material parameters and base thicknesses below 1 µm.

We have calculated the theoretical efficiency of the three structures in the radiative limit applying the detailed balance theory [10]. We have considered variable band-gap energies in the top and middle junctions (emitter and collector in the HBTSCs) and two possible solar spectra: the 6000K black-body spectrum (scaled by $46{,}050^{-1}$ to approximate the irradiance to one-sun level) and the AM15G spectrum. We have assumed that each sub-cell absorbs all the light in its spectral range and does not re-emit light to the cell underneath.

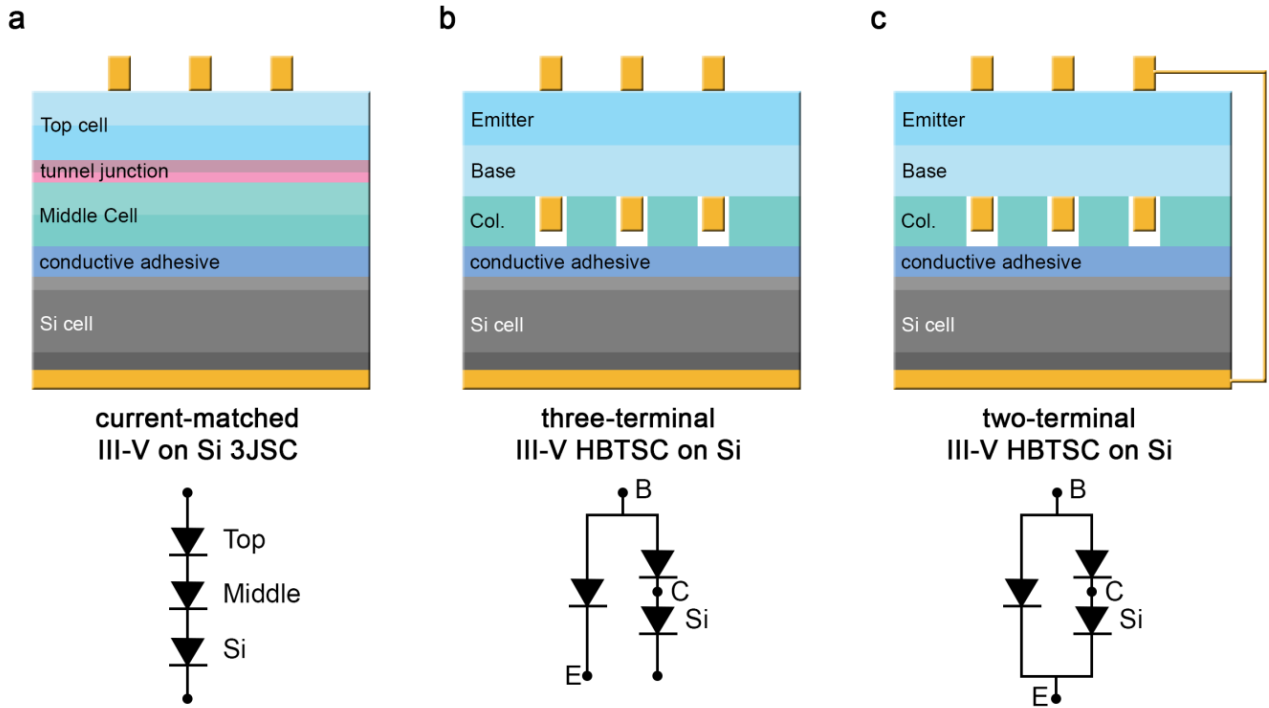

Fig. 2. Device structure of three possible III-V-on-Si triple-junction cells and equivalent circuits.

Results for the one-sun 6000K black-body spectrum are presented in Fig. 3 and for the AM1.5G spectrum in Fig. 4. Both spectra produce similar trends. The efficiencies for AM1.5G are slightly higher, as it is usually found, because this spectrum contains less photons of very high energy, which are inefficiently exploited using typical band-gap energies. Also, high-efficiencies are shifted towards lower band-gap energies, for the same reason. This is especially noticeable for the current-matched AM1.5G.

The highest achievable efficiencies (efficiency limits) for each concept and each spectrum are compiled in Table I, together with the corresponding band-gap energies. The efficiency limit of the current-matched MJSC with Si bottom cell is 0.47 under black-body spectrum, far below the 0.63 absolute efficiency limit of the current-matched 3JSC [11]. The reason is that the band-gap energy of silicon, 1,12 eV, is much larger than the optimum, 0.58 eV. The Si cell produces too little current and limits the other junctions. Under AM1.5G spectrum the efficiency limit raises to 0.49.

For the two HBTSC-on-Si solar cells the efficiency limits are virtually the same than for the current-matched 3JSC: 0.47 with black-body spectrum and 0.49 with AM1.5G (0.50 in the three-terminal configuration). What is remarkable about the HBTSC-on-Si is that the range of band-gap energies for which the efficiency limit is above 0.40 is much wider than in the current-matched MJSC. Moreover, although this range is reduced when the sub-cells are interconnected to produce a two-terminal device, it is still notably large.

That the theoretical efficiency of a three-terminal device is less dependent on the band-gap energies is expected. The reason is that a far-from-optimal band-gap energy in the top junction does not prevent the production of power in the middle and bottom junctions, and vice versa. A similar argument explains why a three-terminal device is more tolerant to spectral changes, and therefore, has higher annual energy yield [12]. The great advantage of the HBTSC-on-Si approach is that this improvement is achieved with a very compact, practical design. Even more so if the two-terminal configuration is chosen. Due to the combination of a series constraint with a parallel constraint, the two-terminal HBTSC-on-Si device is still much more tolerant to non-optimum band-gap energies than the traditional current-matched 3JSC (and, therefore, is expected to be much more tolerant to spectral changes), while it can be integrated in a module with the same ease.

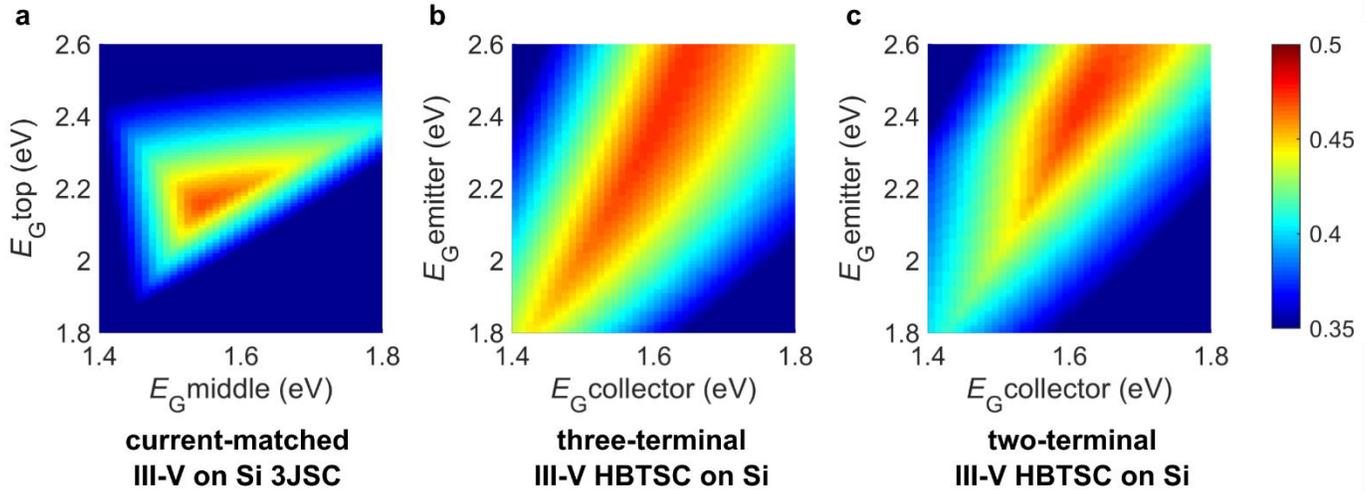

Fig. 3. Detailed balance efficiency limit for the three three-junction configurations depicted in Fig.2, with silicon bottom cell and variable band-gap energies in the upper junctions. The illumination level is one-sun and the spectrum is the 6000K black-body spectrum.

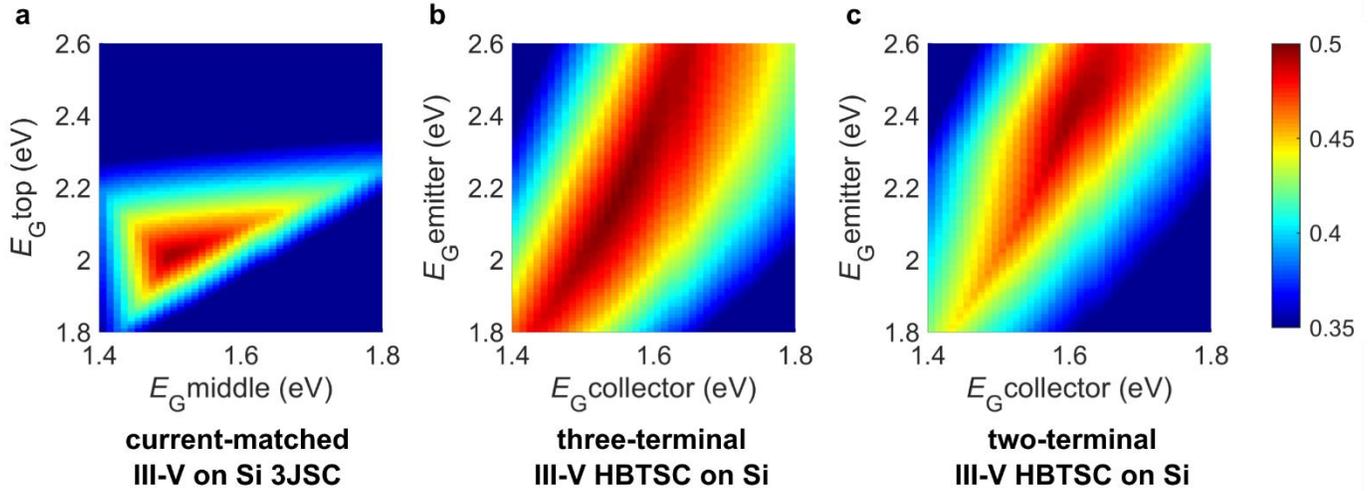

Fig. 4. Detailed balance efficiency limit for the three three-junction configurations depicted in Fig. 2, with silicon bottom cell and variable band-gap energies in the upper junctions. The illumination level is one-sun and the spectrum is AM1.5G.

TABLE I
ONE-SUN DETAILED BALANCE EFFICIENCY LIMITS

|  | Current-matched III-V on Si 3JSC | Three-terminal III-V HBTSC on Si | Two-terminal III-V HBTSC on Si |
|---|---|---|---|
| | One-sun 6000K black-body spectrum | | |
| Maximum efficiency ($E_G$ values) | 0.47 (2.15, 1.54 eV) | 0.47 (2.46, 1.62 eV) | 0.47 (2.45, 1.62 eV) |
| GaInP/GaAs | 0.32 | 0.44 | 0.41 |
| | AM1.5G spectrum | | |
| Maximum efficiency ($E_G$ values) | 0.49 (2.01, 1.50 eV) | 0.50 (2.18, 1.55 eV) | 0.49 (2.41, 1.6 eV) |
| GaInP/GaAs | 0.39 | 0.47 | 0.43 |

IV. THE GaInP/GaAs/Si CASE

Because the HBTSC-on-Si concept is less sensitive to the choice of band-gap energies than the current-matched 3JSC with Si, it is possible to choose the materials that appear more suitable from a technological point of view. A very attractive option is the lattice-matched combination GaInP/GaAs (1.85/1.42 eV). In Table I it can be seen that the AM1.5G detailed balance efficiency with GaInP/GaAs is 0.39 for the current-matched 3JSC, 0.47 for the three-terminal HBTSC-on-Si and 0.43 for the two-terminal HBTSC-on-Si. Lattice-matched GaInP/GaAs are attractive because they can be produced with much better quality (better carrier mobilities and diffusion lengths, more controlled doping) than the complicated quaternary alloys. Nowadays, it is possible to fabricate GaInP and GaAs devices with very high radiative efficiencies [13].

Besides, the fabrication ease usually translates into a lower cost. For example, lattice-matched GaInP and GaAs are highly compatible with the use of high-throughput epitaxial techniques [14]. Also, it is possible that a GaInP/GaAs HBTSC has less substrate-related costs than traditional III-V multi-junctions. The high-quality substrates required for epitaxial growth constitute the primary cost of III-V solar cells and it has been proven that substrate re-use cannot mitigate this problem because of the high cost of preparing the substrate again for each growth [15]. From an economic perspective, it would be more favorable to grow many solar cell structures one on top of another on a single substrate, lift them off one by one and reuse the substrate. The growth of quaternaries and metamorphic layers, required to obtain a particular band-gap energy, and the insertion of highly-doped tunnel junctions are factors that prevent the high-quality growth of sequential MJSCs on a single substrate. Lattice-matched GaInP-GaAs HBTSC structures, with no metamorphic layers and no tunnel junctions, would be excellent candidates for sequential growth, which would result in a very low growth cost per device compared to current III-V technologies.

To understand why the GaInP/GaAs/Si combination works better with the HBTSC than with the traditional current-matched device, we have plotted in Fig. 5 the detailed balance current-density – voltage ($J$-$V$) curve produced by each junction when the stack is illuminated with AM1.5G and the sub-cells are independently connected. It becomes evident that the three cells can only be matched in current with a great loss. The main difference between the current-matched 3JSC and the HBTSC-on-Si is that in the 3JSC the low photocurrent produced in the Si cell limits the current of the GaAs and GaInP cells, whereas in the HBTSC it only limits the production of the GaAs cell. When the sub-cells in the HBTSC-on-Si are interconnected to achieve two terminals, the Si cell is forced to produce a much smaller voltage, but this constraint results in a smaller power loss than the series connection. Interestingly, the Si cell is the only one in the stack that cannot approach the radiative limit in practice. That means that, in practical devices, the Si cell will have the largest difference in $V_{OC}$ with respect to the detailed balance curves. Since the Si cell in the two-terminal HBTSC-on-Si device is working at a low voltage in any case, the efficiency will not drop significantly when non-radiative recombination in the Si cell is taken into account, whereas in the case of the 3JSC the effect will be strong.

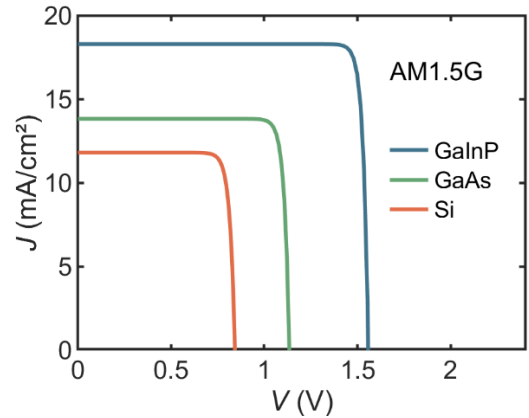

Fig. 5: $J$-$V$ curves calculated for GaInP, GaAs and Si cells with the detailed balance model and the AM1.5G one-sun spectrum. The cells are assumed to be independently connected and to form a stack. Therefore, each cell receives the light that the upper cell(s) lets through.

To attain a more precise understanding of the efficiency limits calculated here, it will be interesting to introduce in future calculations the non-radiate recombination in the Si cell. Also, the possible optical thinning of the III-V sub-cells, which could alleviate to some extent the power loss produced by the sub-cell constraints. If the III-V materials are highly radiative, the effect of luminescent coupling could be similar to the one achieved with optical thinning [16].

V. CONCLUSIONS

We have proposed a III-V-HBTSC–on-Si solar cell. Detailed balance calculations show that, both for the three-terminal and the two-terminal configuration, the one-sun efficiency limit is above 0.40 for a great range of III-V material band-gap energies. In particular, the AM1.5G detailed balance efficiency for a GaInP/GaAs/Si device is 0.47 with a three-terminal and 0.43 with a two-terminal configuration. The combination GaInP/GaAs has many practical advantages with respect to the III-V quaternary compounds that produce similar efficiencies in a current-matched MJSC. Our results show that the III-V-HBTSC-on-Si device structure has the potential for developing a low-cost, high-efficiency PV technology.